\documentclass[traditabstract]{aa}

\usepackage{graphicx}
\usepackage{rotating,subfigure,amssymb,afterpage}
\usepackage{txfonts}
\usepackage{natbib}
\usepackage[pdftitle={Local Underdensity in the{\sf REFLEX II} Survey},
pdfauthor={H. B\"ohringer et al.},
pdfsubject={}, pdfkeywords={}, bookmarks, bookmarksopen, 
colorlinks=true, linkcolor=blue, citecolor=blue, anchorcolor=blue,
urlcolor=blue]{hyperref} 
\newcommand{\propsim}{\lower 3pt \hbox{$\, \buildrel {\textstyle
      \propto}\over {\textstyle \sim}\,$}}
\begin{document}
   \title{The extended ROSAT-ESO Flux-Limited X-ray Galaxy Cluster
   Survey (REFLEX II)\\ VII The Mass Function of Galaxy Clusters}

      \author{Hans B\"ohringer\inst{1}, Gayoung Chon\inst{1}, Masataka Fukugita\inst{2}} 

   \offprints{H. B\"ohringer, hxb@mpe.mpg.de}

   \institute{$^1$ Max-Planck-Institut f\"ur extraterrestrische Physik,
                   D-85748 Garching, Germany.\\
              $^{2}${Kavli Institute for the Physics and Mathematics of the Universe,
                   University of Tokyo, Kashiwa 2778583, Japan}}

   \date{Submitted 19/5/17}

\abstract{The mass function of galaxy clusters is a sensitive 
tracer of the gravitational evolution of 
the cosmic large-scale structure and serves as an important
census of the fraction of matter bound in large structures.
We obtain the mass function by fitting the observed cluster X-ray
luminosity distribution from the {\sf REFLEX} galaxy cluster survey
to models of cosmological structure formation.
We marginalise over uncertainties in the cosmological 
parameters as well as those of the relevant galaxy cluster scaling relations.
The mass function is determined with an uncertainty less than 10\% in the 
mass range $3 \times 10^{12}$ to $5\times 10^{14}$ M$_{\odot}$. 
For the cumulative mass function we find a slope at the low mass 
end consistent with a value of $-1$,
while the mass rich end cut-off is milder than a Schechter function with
an exponential term $exp(- M^{\delta}$) with $\delta$ smaller than 1.
Changing the Hubble 
parameter in the range $H_0 = 67 - 73$ km s$^{-1}$ Mpc$^{-1}$ or allowing
the total neutrino mass to have a value between $0 - 0.4$ eV
causes variations less than the uncertainties.
We estimate the fraction of mass locked up in galaxy clusters:
about 4.4\% of the matter in the Universe is bound
in clusters (inside $r_{200}$) with a mass larger than $10^{14}$ M$_{\odot}$
and 14\% to clusters and groups with a mass larger than $10^{13}$ M$_{\odot}$
at the present Universe.
We also discuss the evolution of the galaxy cluster population with redshift.
Our results imply that there is hardly any clusters with a mass 
$ \ge 10^{15}$ M$_{\odot}$ above a redshift of $z = 1$.
}  

\authorrunning{B\"ohringer et al.}
\titlerunning{{\sf REFLEX II} the galaxy cluster mass function}
   \maketitle
%

\section{Introduction}

Clusters of galaxies form from peaks in the density fluctuation 
field in the Universe in a well prescribed way. The statistics of
the peaks is clearly characterised for a given cosmological model.
Therefore in the framework of the currently well established 
cosmic structure formation theory, the abundance of galaxy clusters 
as a function of their mass can be predicted 
(e.g. Press \& Schechter 1974, Bardeen et al. 1986, Bond et al. 1991,
Jenkins et al. 2001, Warren et al. 2006, Tinker et al. 2008, Despali
et al. 2016). This cluster mass function 
is a very sensitive tracer of the  growth of structure in
the Universe and its observational determination can be applied
to stringent tests of cosmological models. In most cases in
these cosmological tests the mass function is substituted by a
function of another more easily observed cluster property which can
be used as a good mass proxy, like cluster richness, X-ray luminosity
or temperature, or the Sunyaev-Zel'dovic effect decrement (e.g.
Perenod 1980, Reiprich \& B\"ohringer 2002, Henry 2004
Vikhlinin et al. 2009, Rozo et al. 2010, Mantz et al. 2010, 
B\"ohringer et al. 2011, Planck Collaboration 2014, 2016,
B\"ohringer et al. 2014, B\"ohringer \& Chon 2015).

Since it is very useful to know the cluster mass function for 
many applications, apart from the testing of cosmological models,
as for example for models of the halo occupation distribution (e.g. 
Peacock \& Smith 2000, Berlind \& Weinberg 2002, Mehrtens et al. 2016) 
it is important to compile our knowledge on the mass function
explicitly. For this purpose we use our X-ray cluster survey based on
the ROSAT All-Sky Survey, which is currently the largest, statistically
most complete and well described cluster survey in the nearby Universe
covering most of the sky outside the galactic band.
We focus here on the {\sf REFLEX II} cluster survey in the southern sky, 
because in the northern counterpart, the {\sf NORAS II} survey, there are 
still some cluster candidates with missing redshifts (B\"ohringer
et al. 2017).

One conventional way to derive the mass function from observations
is to use the X-ray luminosity distribution and convert the data via 
an empirical X-ray luminosity - mass scaling relation. We like 
to explore a different approach.
Since the theory of structure formation is well established and has 
been tested against many observational results, we have a good theoretical
background knowledge on the mass function and its relation
to the cosmological model. We can use this as additional information
to constrain the mass function by requesting that it is not only consistent
with the data but also with structure formation theory. 
In practice, this type of mass function appears as a by-product 
when we test cosmological models, and does not require additional
calculations. The method is described in detail in section 2. 
It provides very tight constraints on the mass function.

The paper is structured as follows. 
Section 2 provides a brief account of the {\sf REFLEX} cluster survey
and its use for cosmological tests. Section 3 describes the method
to derive the mass function. In section 4 we present the 
results and compare with previous observations. We look at further 
implications of our results in section 5 including the fraction of
cosmic matter bound in clusters, the implied redshift evolution 
of the mass function, and the predicted total number counts
of clusters. Section 6 then provides a summary and conclusions.
We quote and display all our results scaled by a Hubble parameter of
$h_{70} = H_0/70~ {\rm km~ s}^{-1}~ {\rm Mpc}^{-1}$.

\section{The REFLEX II Galaxy Cluster Survey}

The {\sf REFLEX II} galaxy cluster survey produced a catalog of 911
X-ray luminous galaxy clusters down to a flux limit of
$1.8 \times 10^{-12}$ erg s$^{-1}$ cm$^{-2}$ in the 0.1 - 2.4 keV 
energy band with an estimated completeness of about 95\%
and a possible significant X-ray contamination by AGN
in about 6\% of the sources
(B\"ohringer et al. 2013). 
It is based on the X-ray detection of the
clusters in the ROSAT All-Sky Survey (Tr\"umper 1993, 
Voges et al. 1999) and covers a region in
the southern sky below equatorial latitude +2.5$^o$ and at galactic
latitude $|b_{II}| \ge 20^o$. The regions of the Magellanic clouds have
been excised. The total survey area is $ \sim 4.24$ ster.
The source detection, the
galaxy cluster sample definition and compilation, and the construction of
the survey selection function  as well as tests of the completeness of the
survey are described in B\"ohringer et al. (2013). 

The X-ray fluxes and derived luminosities have been determined
by means of the growth curve analysis method (B\"ohringer et al. 2000)
within a fiducial cluster radius of 
$r_{500}$ \footnote{$r_{500}$ is the radius where the average
mass density inside reaches a value of 500 times the critical density
of the Universe at the epoch of observation}. $r_{500}$ is derived
from the mass of the clusters, which is obtained by
the X-ray luminosity - mass relation (B\"ohringer et al. 2014, Eq. 10)
obtained by Pratt et al. (2009)
also consistent with that of Vikhlinin et al. (2009).
\footnote{While most of the X-ray observational data are
limited to the region inside $r_{500}$, we use $M_{200}$
as cluster mass parameter for better comparison with the literature.
The conversion from $M_{500}$ to $M_{200}$ is performed by an 
extrapolation based on the well established NFW profile
(Navarro, Frenk \& White 1995).}
The uncertainty of the flux and luminosity measurement is on
average about 20\%.

In B\"ohringer et al. (2014, 2016) and B\"ohringer \& Chon (2015) 
we used the {\sf REFLEX} cluster sample to test cosmological models
by comparing the observed distribution of cluster X-ray luminosities
in the cluster catalog from the {\sf REFLEX II} project
with model predictions.
The comparison is performed by means of a likelihood method
described in detail in (B\"ohringer et al. 2014).
In these calculations the model for structure formation is 
based on a primordial power spectrum of the matter density
fluctuations with a slope of 0.96 and a transfer function
which models the evolution of the power spectrum obtained 
with the program CAMB (Lewis et al. 2000)
\footnote{ CAMB is publicly available from
http://www.camb.info/CAMBsubmit.html}. The prediction of the cluster
mass function from the matter power spectrum is performed with
the formulas given by 
Tinker et al. (2008). Specifically we have used
their Eqs. 2 and 3 with the values $A = 0.186$, $a = 1.47$,
$b = 2.57$, and $c = 1.19$ for the redshift zero mass function
and applied their Eqs. 5 to 7 for the redshift evolution.
To derive the predicted X-ray 
luminosity function from the theoretically calculated cluster
mass function, we used empirical scaling relations of 
X-ray luminosity and mass (B\"ohringer et al. 2014, eq. 10)
in accordance with the observations
of Reiprich \& B\"ohringer (2002), Pratt et al. (2009) and
Vikhlinin et al. (2009), within their confidence limits.

The predicted X-ray luminosity function depends most strongly
on the two cosmological parameters, $\Omega_m$ and $\sigma_8$.
Thus these are the parameters to be best constrained with
the galaxy cluster data. The constraining power of the
data is mostly limited by systematics, that is by our 
imprecise knowledge of the cluster scaling relation of
X-ray luminosity and mass (e.g. B\"ohringer et al. 2012).
To derive realistic constraints on $\Omega_m$ and $\sigma_8$
we therefore marginalise over the uncertainties in the two most 
influential and important parameters: we allow for a 
7\% uncertainty in the slope of the scaling relation and an
uncertainty of 14\% in its normalisation (equivalent to the mass 
calibration) as 1$\sigma$ constraints of these parameters.
We also include a scatter of the scaling relation of 30\%
and in addition a mass bias of 10\%
(B\"ohringer et al. 2014). The latter acounts for the fact
that the cluster masses in our empirical scaling relations,
which have been determined from X-ray observations assuming
hydrostatic equilibrium of the intracluster medium, are on
average biased low. 
For more details on the marginalisation see B\"ohringer et al. 
(2014). The result of the likelihood analysis including this
marginalisation is shown in Fig. 1
\footnote{The uncertainty in the theoretical modelling 
of the mass function has not been considered in the above error budget.
It corresponds to a systematical uncertainty of a few percent, typically
not more than 5\% for $\Omega_m$ and $\sigma_8$.} . It implies the following
results, $\Omega_m = 0.285 \pm 0.04$ and $\sigma_8 =  0.776 \pm 0.07$.
  
\section{Method}

\begin{figure}
   \includegraphics[width=\columnwidth]{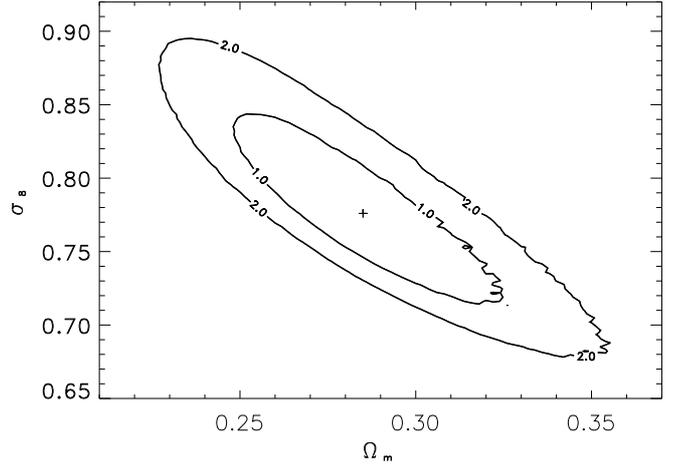}
\caption{Constraints on the cosmological parameters $\Omega_m$
and $\sigma_8$ from the {\sf REFLEX} galaxy cluster survey
obtained by marginalising over the uncertainties in the cluster
scaling relation for X-ray luminosity and mass (from B\"ohringer
et al. 2014). The cross shows the best fitting parameters and the
ellipses indicate 1 and 2$\sigma$ uncertainties.
}\label{fig1}
\end{figure}

In the above analysis the mass function appears as an intermediate
product. It is calculated for a given cosmological model, and subsequently
used to predict the X-ray luminosity distribution of the clusters
which is fitted to the observational
data. The nominal mass function is that one which corresponds to the
cosmological model fitting best to the observational data. To determine
the uncertainty of the mass function we use the marginalisation
approach described above. We sample the distribution of mass functions
corresponding to the parameter distribution shown in Fig. \ref{fig1}.
For each pair of cosmological parameters there is also a best fitting
set of scaling relation parameters \footnote{Including the variation
of the scaling relation parameters is necessary for the following
reason. For a fixed set of scaling parameters the error ellipse
for $\Omega_m$ and $\sigma_8$ is much smaller than shown in Fig. 1
(see Fig. 9 in B\"ohringer et al. 2014). For the full parameter range
allowed by the constraint limits in Fig. 1, good fits to the mass 
function are only provided for all values of $\Omega_m$ and $\sigma_8$
if the scaling relation parameters are also optimised. Therefore
the mass function is only consistent with the data for certain
combinations of cosmological and scaling relation parameters.}.   
To determine the uncertainty of
the mass function consistent with the cosmological model framework,
we sample 100 parameter sets from statistical distribution shown
in Fig. \ref{fig1}, and calculate 100 different mass functions
accordingly. For this set of functions 
we determine the 16\% and 84\% percentiles as $1\sigma$ limits.

We consider cosmologies without massive neutrinos and a
model with a sum of the neutrino masses of $\sum m_{\nu} = 0.4$ eV.
We also study the influence of the variation of the Hubble parameter in
the range from  $H_0 = 67$ to $73$ km s$^{-1}$ Mpc$^{-1}$.

\section{Results}

\begin{figure}
   \includegraphics[width=\columnwidth]{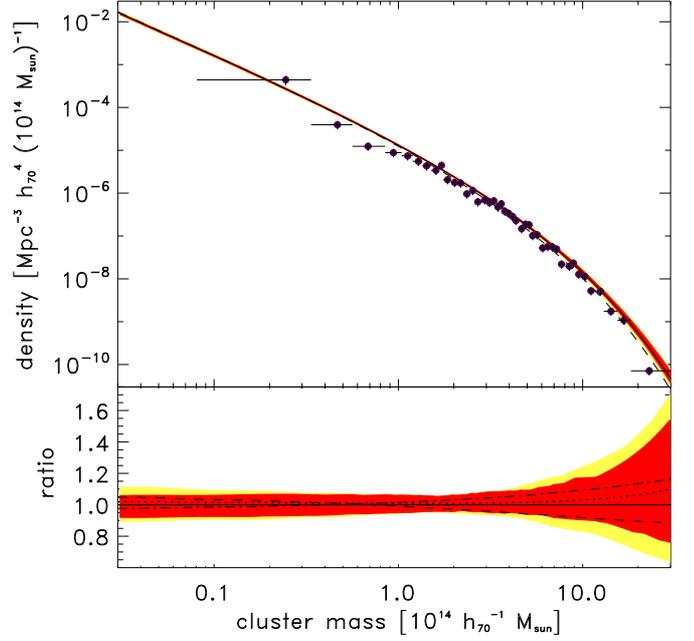}
\caption{{\bf Upper panel:} 
Mass function for the cosmological model best fitting the
{\sf REFLEX} Survey data (thick black line).
The dark shaded region shows
the uncertainties while marginalising over the 1$\sigma$ limits of 
the most important cosmological and scaling relation parameters
(see text). 
The lightly shaded region shows in addition the variation with 
the Hubble constant in the range $67$ to $73$ km s$^{-1}$ Mpc$^{-1}$.
The data points give the mass function derived from the 
X-ray luminosity function of the {\sf REFLEX} cluster survey.
The dashed line gives the best fitting mass function for a 
redshift of $z = 0.102$ for comparison with the {\sf REFLEX} data.
{\bf Lower panel:} ratio of the uncertainty limits to the best
fitting function, where the light and dark shaded regions have the
same meaning as above. The dashed and dashed-dotted lines show the ratio for the best
fitting functions for a Hubble constant of $67$ and $73$ km 
s$^{-1}$ Mpc$^{-1}$. The dotted line shows the best fitting function
allowing the sum of the mass of the neutrinos to be $0.4$ eV,  
}\label{fig2}
\end{figure}

Fig. \ref{fig2}  shows the results for the mass function implied by the best
fitting cosmological model, the 1$\sigma$ uncertainties from
the marginalisation and the uncertainties from allowing an additional
flat prior on $H_0$ in the range between $ 67$ to $73$ km s$^{-1}$ Mpc$^{-1}$.
The cluster masses were calculated for a fiducial radius of $r_{200}$, 
the radius inside which the mean matter density is 200 times 
the critical density at the cluster redshift. 
Note, that the mass function shown is what is implied for a redshift
of zero, while the mass function used for the fit to the {\sf REFLEX} 
cluster data is for a redshift of $z=0.102$.
The cluster mass function is tightly constrained with an uncertainty
less than 10\% in the mass range $ 3\times 10^{12}$ to 
$5\times 10^{14}$ M$_{\odot}$. The function is less well 
constrained at the high mass end. The uncertainties are displayed in
more detail in the lower part of the plot,
which shows the ratio of the upper and lower limits to the best fitting 
function. We note that the variation of the Hubble constant 
in the above quoted range changes the results by an amount smaller
than the uncertainties.

We compare these results of the mass function, to data points of
the {\sf REFLEX} sample in the Figure. These data have been obtained
by converting the X-ray luminosity function from (B\"ohringer et al. 2014)
to a mass function by use of the $L_X - M$ scaling relation.
This is over-plotted in Fig. \ref{fig2} with data points for the binned
mass function of the {\sf REFLEX} sample with 20 clusters per bin, except
for the bin at the lowest masses, which has only three clusters. Since the data 
correspond to a median redshift of $z = 0.102$ we also plotted in
Fig. \ref{fig2} the mass function predicted for this redshift
as dashed line. This line fits the observational data very well.
One can observe a dip of two to three data points at the low mass end,
which deviates from the line. This is mostly an effect of cosmic
variance, caused by the matter underdensity in the nearby Universe
in the southern sky, detected by our {\sf REFLEX II} survey
(B\"ohriner et al. 2015). Thus we conclude that the two derivations 
of the galaxy cluster mass function are consistent. We 
also note, that our new approach, including the consistency
with sructure formation theory, provides tighter constraints.

The results shown have been derived for a negligible mass 
of all neutrino families. Allowing for a minimum mass of neutrinos
given by the constraints from particle physics of 
$\sum m_{\nu} = 0.06$ eV has a 
hardly noticible effect on the results. To better
demonstrate the effect of a neutrino mass we have taken 
a larger value of  $\sum m_{\nu} = 0.4$ eV (motivated also by the
upper limit of our
own cosmological constraints obtained with the  {\sf REFLEX II} survey
in B\"ohringer \& Chon 2015). We find that the change is smaller than the 
uncertainties (as long as no further external observational data are included). 
This is shown in the lower panel of Fig. \ref{fig2}.

\begin{figure}
   \includegraphics[width=\columnwidth]{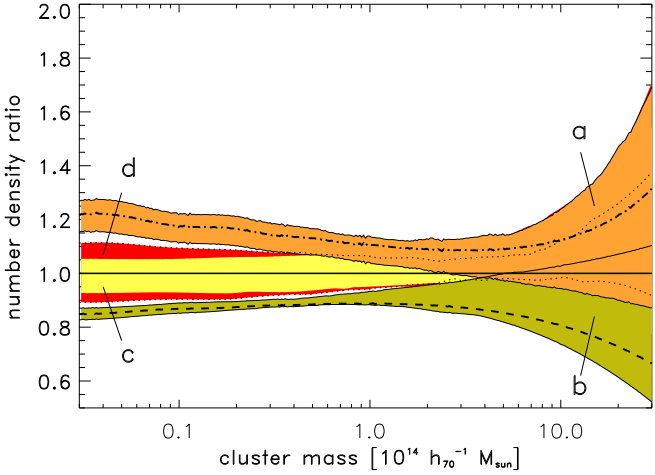}
\caption{Galaxy cluster mass function for the cosmological model best
fitting the {\sf REFLEX} Cluster Survey data based on different cluster 
mass function models. 
The mass functions are normalised to the results based on the Tinker 
mass function (solid line). The dashed line shows the results for the
mass function model by Watson et al. (2013) with the uncertainty range
labeled by (b) and the dashed-dotted line
the results by using the model by Despali et al. (2016) 
with uncertainties labeled by (a). 
The yellow-shaded region labeled (c)
shows the uncertainties after marginalisation for the mass function
based on Tinker et al. and the red region labeled (d) indicates the additional
uncertainties from the marginalisation over the uncertainty
in the Hubble parameter.
}\label{fig3}
\end{figure}

The  tight constraints of the cluster mass function rely, however,
on the precision of the mass function model by Tinker et al. (2008) 
used in our analysis. Therefore it is useful to explore how the function
changes with the use of other published mass function models.
In Fig. \ref{fig3} we provide the results for the mass
function based on models proposed by Watson et al. 
(2013) and Despali et al. (2016). The constraints shown for the
use of Watson's algorithm are based on the
model mass function for AHF halos defined as overdensities.
\footnote{Watson et al. (2013) provide three different mass function 
models: one based on a friends-of-friends (FoF) halo finder, the AHF model
defined by an overdensity threshold, and the CMPSO algorithm also
based on sperical overdensity. Using the different models results in 
very similar mass function constraints with differences barely larger
than the uncertainties. The largest difference is on the large mass end
above $10^{15}$ M$_{\odot}$, where the FoF method
predicts more clusters than Tinker et al. (2008) and the spherical overdensity
methods stay on average below.}
The results show reasonable 
agreement, with variations of up to 20\% at the outer extremes of
the displayed mass range. The function of Watson et al. predicts a 
smaller number of less massive clusters, while the model of Despali et al. 
predicts a slightly higher number of clusters over the whole mass range 
than Tinker et al. (2008).

We note that, as a consequence of the difference in the best 
fitting cosmological parameters between {\sf Planck} CMB and clusters,
the derived mass function is significantly lower,
than what is predicted by using cosmological parameters in
accord with the published results from the {\sf Planck} mission
(Planck Collaboration 2014, 2016). At the high mass end the cluster
abundance is about a factor of two higher for {\sf Planck} cosmology
than for the mass function
derived here. This has been illustrated in our earlier publications
(B\"ohringer et al. 2014, B\"ohringer \& Chon 2015, 2016 -- see in particular
Fig. 11 of B\"ohringer et al. 2014) and reasons for this discrepancy
have been discussed. 

\begin{figure}
   \includegraphics[width=\columnwidth]{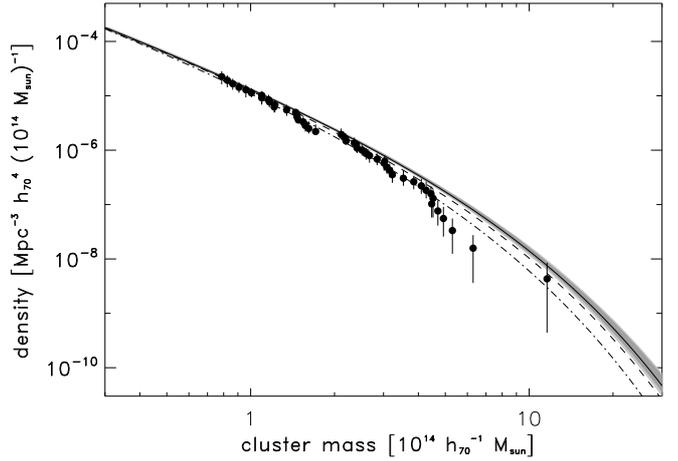}
\caption{Comparison of the observationally constraint mass function
for redshifts of $z = 0$, $z = 0.1$, and $z = 0.25$ with observational 
results by Vikhlinin et al. (2009) for a flux-limited sample of 49
clusters in the redshift range $z = 0$ to $0.25$.
}\label{fig4}
\end{figure}

An accurate empirical mass function of galaxy clusters was also
derived by Vikhlinin et al. (2009) from detailed Chandra 
observations of a flux limited sample of 49 nearby galaxy clusters.
The total mass of these clusters has been derived on one hand
from good measurements of the total gas mass in the clusters 
and a calibration of the gas to total mass ratio from simulations and
on the other hand from observations of the total thermal energy
in the clusters (gas mass times temperature) and the simulation calibrated
scaling relation of this parameter with the total mass
(Vikhlinin et al. 2009). Since the clusters of the Vikhlinin sample cover a redshift 
range from $z = 0$ to $0.25$ we compare the mass function data points
in Fig. \ref{fig4} to the observationally constraint model predicted mass function 
for $z = 0$, $z = 0.1$, and $z = 0.25$. As the publication by Vikhlinin 
et al. is based on a Hubble constant different from the one used here,
we rescaled their data accordingly.
We see that, apart from five data points
at high mass, the model mass functions bracket the data very well.
We actually observe, that at low mass the data points are close to the  
$z =0$ function and at high mass they approach the $z = 0.25$ function
as expected for a flux limited sample, with low mass objects very
nearby and high mass objects only sampled by the larger volume 
including higher reshifts. 

\begin{figure}
   \includegraphics[width=\columnwidth]{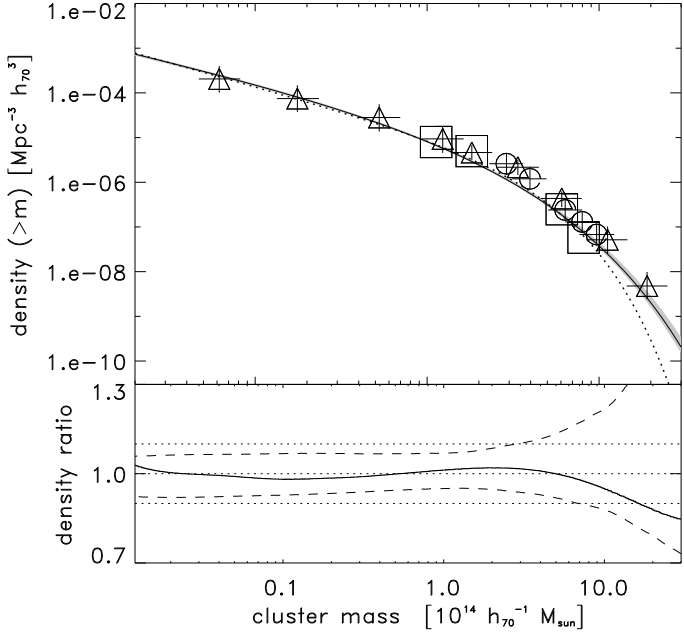}
\caption{{\bf Upper panel:} Cumulative cluster mass function for 
the cosmological model best fitting the {\sf REFLEX} Survey data
and its uncertainties (grey region). The present data are
compared to a previous determination of the cumulative mass function
of clusters by Bahcall \& Cen (1993), shown by data points. 
The different data points relate to different cluster
samples used by Bahcall \& Cen: Abell clusters (triangles with error bars),
EDCC clusters (squares), and X-ray clusters (circles with error bars).
The dotted curve shows the best fitting Schechter function, which 
falls short in describing the high end of the mass function.
{\bf Bottom panel:} Ratio of the fitted function to the true cumulative mass
function. We also show the uncertainties of the mass function
from the marginalisation not including the varying Hubble parameter
(dashed lines). The dotted lines indicate 10\% deviations. 
}\label{fig5}
\end{figure}

In Fig. \ref{fig5} we derive the cumulative mass function,
the cluster density for all clusters above a lower mass limit.
This function was previously determined for example
by  Bahcall \& Cen (1993).
We compare their results to our constraints in the Figure,
where we have indicated the different cluster samples of these
authors coming from a complete sample of Abell clusters 
(Abell 1958), from optical data for the EDCC clusters 
(Lumsden et al. 1992, Nichol et al. 1992), and X-ray clusters
from Henry \& Arnaud (1991) by different symbols. Although
the cluster number density is slightly overestimated in
these data in the middle of the mass range, these early
results are in remarkable agreement with our derivation. 

While the mass function for {\sf REFLEX} has been constructed for
a mass of $M_{200}$, the masses by Bahcall \& Cen were originally
determined for a radius of $ 1.5 h_{100}^{-1}$ Mpc and different
cosmological parameters. To make the comparison we have converted
their masses to our fiducial cosmological model and we have 
also translated the mass results to a radius of $r_{200}$
by means of the NFW mass profile (Navarro, Frank \& White 1996) 
with a concentration parameter of 5.

\section{Application}

\subsection{Fits to the cumulative mass function}

For an easy application of our results to cosmological 
modeling, we provide an analytic fit to the derived 
cumulative mass function at z = 0. A Schechter function 
provides an unsatisfactory fit to the data, the function 
decreases too fast at high mass. We found the 
following function with four free parameters to provide 
a better fit:

\begin{equation} 
n(>M) =  \alpha~ \left( {M \over 2 \times 10^{14} h_{70}^{-1} {\rm M}_{\odot} }\right)^{-\beta}
 ̃̃exp\left(- {M^{\delta} \over \gamma}\right) ~~~~~~.
\end{equation} 

Fig. \ref{fig5} shows the deviation of this fit from the observationally
derived function. Within a mass range 
of $3 \times 10^{12} -  10^{15} h_{70}^{-1}$ M$_{\odot}$ the fit reproduces
the mass function with a error less than 5\%. In the 
Figure we also show the uncertainty of the mass function 
obtained by the marginalsation over the important 
cosmological and scaling parameters not including the 
variations with $H_0$. The deviations of the fitted functions 
are smaller than the uncertainties.

In Table 1 we provide the parameters for the best fit, for
the lower, and upper 1$\sigma$ limit without and with marginalisation
over the Hubble parameter. We also give the fit to the cumulative
mass function obtained for a cosmology with a total neutrino
mass of $ \sum_i m_{\nu_i} = 0.4$ eV. In addition we provide a fit for the best
fitting mass function with the parameter $\beta$ in Eq. (1) fixed
to a value of -1. This three parameter fit is also an acceptable representation
with less than 10\% deviation in the relevant mass range.  

   \begin{table}
      \caption{Fit parameters for the approximations to the cumulative mass function}
         \label{Tempx}
      \[
         \begin{array}{lrrrrr}
            \hline
            \noalign{\smallskip}
 {\rm function}& \alpha^1  & \beta  & \gamma^2 & \delta \\
            \noalign{\smallskip}
            \hline
            \noalign{\smallskip}
{\rm best~ fit}  & 1.237  & 0.907  & 0.961 & 0.625 \\
{\rm fit~ to~ lower~ limit}  & 1.205  & 0.895  & 0.941 & 0.627 \\
{\rm fit~ to~ upper~ limit}  & 1.231  & 0.922  & 1.026 & 0.629 \\
{\rm fit~ lower~ limit~ incl.~ H_0~ var.}  & 1.252  & 0.980  & 0.907 & 0.625 \\
{\rm fit~ upper~ limit~ incl.~ H_0~ var.}  & 1.163  & 0.942  & 1.083 & 0.636 \\
{\rm fit~ incl.~ massive~ neutrinos}  & 1.367  & 0.904  & 0.895 & 0.608 \\
{\rm best~ fit, 3~ free~ parameters} & 0.821 & 1.000 & 1.372  & 0.710 \\ 
            \noalign{\smallskip}   
            \hline
         \end{array}
      \]
\begin{list}{}{}
\item[$^1${\rm $\alpha$ in units of $10^{-5}$ Mpc$^{-3} h_{70}^3 10^{14}$ M$_{\odot}$}],
\item[$^2${$\gamma$ in units of $\left(10^{14} {\rm M}_{\odot}\right)^{\delta}$}].
\end{list}
\label{tab1}
   \end{table}
%

\subsection{Cosmic mass fraction in cluster and groups}

\begin{figure}
   \includegraphics[width=\columnwidth]{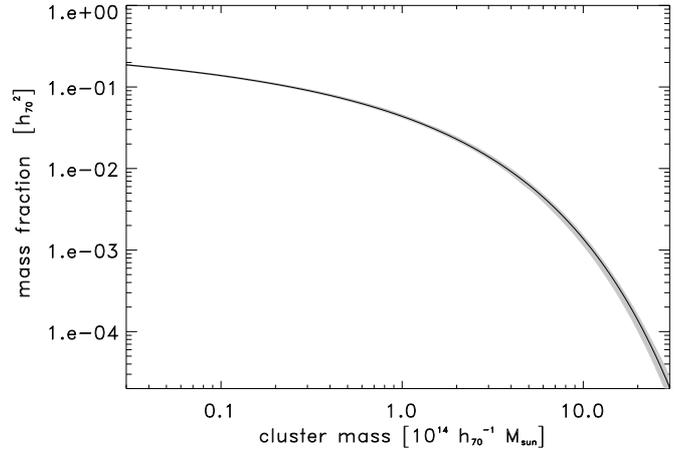}
\caption{Mass fraction bound in groups and clusters of galaxies
inside $r_{200}$ at a redshift of zero.
}\label{fig6}
\end{figure}

The cumulative mass function allows us to calculate the fraction
of matter in the Universe contained in clusters and groups
above a certain mass limit inside a radius of $r_{200}$. This
is shown in Fig. \ref{fig6}, where the mass fraction in groups and clusters
has been normalised by the total amount of matter in the Universe
for $H_0 = 70$ km s$^{-1}$ Mpc$^{-1}$ and the relevant value of $\Omega_m$
which is $0.285$ for the best fitting model. At present about $14 \pm 1\%$
of the matter in the Universe is contained in groups and clusters
with a mass above $10^{13} h_{70}^{-1}$ M$_{\odot}$, and $4.4 \pm 0.4\%$ is locked
up in clusters with a mass above $10^{14} h_{70}^{-1}$ M$_{\odot}$ inside a 
radius of $r_{200}$. If we extend the calculations to higher redshifts,
we find only of the order of 1\% of the matter is bound groups and clusters with
$M_{200} > 10^{13} h_{70}^{-1}$ M$_{\odot}$ at redshift 2, illustrating the recent
rapid evolution of the cluster population.

\subsection{Evolution of the mass function with redshift}

\begin{figure}
   \includegraphics[width=\columnwidth]{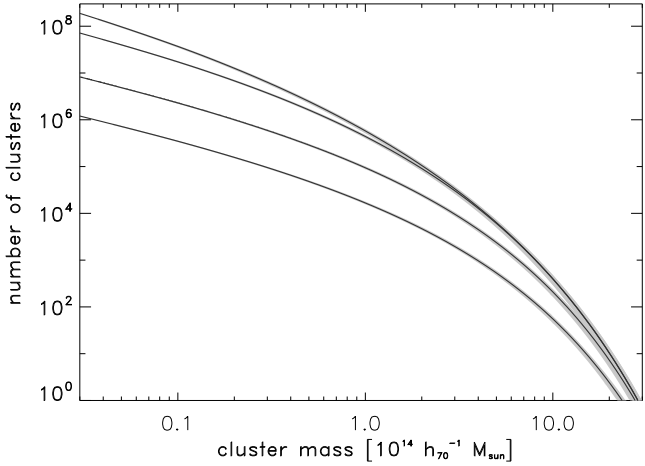}
\caption{Cumulative number counts of clusters in the full sky as a function of
mass for different limiting redshifts. The curves
shows the cut-off redshifts of z = 0.2, 0.4, 1.0, 2.0, respectively.
}\label{fig7}
\end{figure}

We can use our constraints on the evolution of structure formation
to derive further constraints on the cluster population as a function
of redshift. Fig. \ref{fig7} shows the cumulative number
of clusters that can be observed out to a maximum redshift as a function
of the lower cluster mass limit. The total number of groups and clusters 
out to a redshift of 2 is for example predicted to be about $36 \pm 2.5$ 
million with a mass above $M_{200} > 10^{13} h_{70}^{-1}$~M$_{\odot}$ and for clusters 
with a mass above $M_{200} > 10^{14} h_{70}^{-1} $ M$_{\odot}$ about
$560 000 \pm 60 000$ .

\begin{figure}
   \includegraphics[width=\columnwidth]{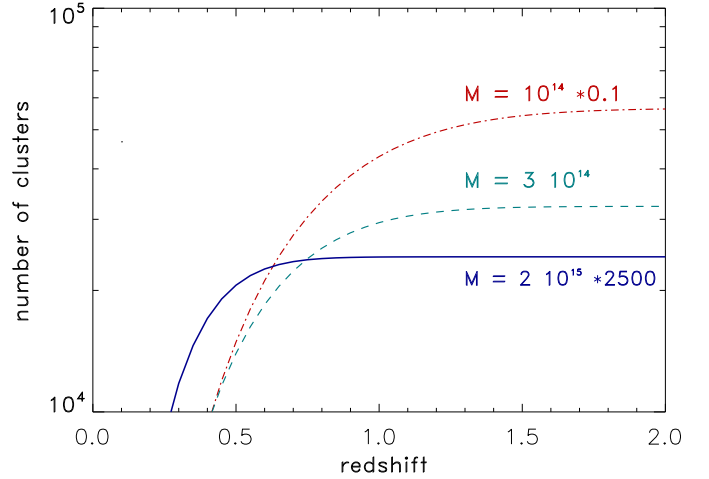}
\caption{Reshift evolution of the cumulative number counts of clusters
for different minimum mass. The lower mass cut-off for the curves are
from top to bottom $10^{14} h_{70}^{-1}$, $3\times 10^{14} h_{70}^{-1}$, and 
$2\times 10^{15} h_{70}^{-1}$ M$_{\odot}$.
}\label{fig8}
\end{figure}

We also note in Fig. \ref{fig7} that the increase in number counts 
at higher redshifts only involves the lighter systems. Thus for more
massive systems, there is practically a limiting redshift, beyond
which we cannot expect to find further objects of this size.
This is further detailed in Fig. \ref{fig8}. We show here the increase
in number counts for systems with $M_{200} > 2 \times 10^{15}$ M$_{\odot}$,
$ > 3 \times 10^{14} h_{70}^{-1}$ M$_{\odot}$, and $ > 10^{14} h_{70}^{-1}$ M$_{\odot}$.
We thus expect to find only about 9 - 10 clusters with a mass above 
$2 \times 10^{15} h_{70}^{-1}$ M$_{\odot}$ and they should all reside at redshifts
lower than $0.75$. In the whole sky we should find about 390 
clusters with $M > 1 \times 10^{15} h_{70}^{-1}$ M$_{\odot}$
with the most distant four clusters  
found in the redshift interval $z = 1.0 - 1.2$.

We can compare this to the actual numbers found in our survey.
For {\sf REFLEX II} and {\sf NORAS II} together
(which we now call {\sf CLASSIX} survey), which cover about
two thirds of the sky (8.25 ster), we find 16 clusters 
with a scaling relation determined
mass $> 2 \times 10^{15} h_{70}^{-1}$ M$_{\odot}$. That this
number is larger than the expected value quoted above results from the 
fact that the scaling relations have a considerable scatter which strongly 
affects the number counts at the high mass end, where the mass function is 
particularly steep. If we include a 30\% scatter in the X-ray luminosity
mass relation, we predict to find 19 high mass clusters, of which 
13 should lay in the {\sf CLASSIX} survey
\footnote{Cosmic Large-Scale Structure in X-rays
Cluster Survey (B\"ohringer et al. 2016)} region. 
For clusters with $M_{200} > 10^{15} h_{70}^{-1}$ M$_{\odot}$ we predict about 
200 such objects for $z < 0.4$. Including a 30\% scatter and accounting for
the sky coverage of {\sf CLASSIX} we predict 183 clusters and observe
161. Thus, taking into account the scatter in the scaling relations
with the same amount as it was used for the cosmological modelling, we get
a good agreement with the predictions and the observations.

\section{Summary and Conclusion}

A major goal of this paper is to provide the mass function
of galaxy clusters as a census of the galaxy cluster population in
the low redshift Universe. We have shown that combining our catalog of X-ray
luminous clusters from the {\sf REFLEX} survey with structure formation
theory, we obtain a mass function with small uncertainties
not larger than 10\% in the mass range $3 \times 10^{12} h_{70}^{-1}$ to
$5\times 10^{14} h_{70}^{-1}$ M$_{\odot}$. We also provided approximations by
parameterised functions, which capture the results well within their 
uncertainties. 

The largest uncertainties included in these results come from the
limited precision of the mass - observable scaling relations of galaxy
clusters, which is also the major limit factor in modelling
galaxy cluster cosmology. Large efforts are currently made to improve
this situation and we hope that this will enable us to derive
a more precise cluster mass function and its evolution
in the future. We also find that there is a significant uncertainty
on the theoretical side with the mass function model. 
If we want to improve the results well below a 10\% accuracy,
also the modelling of the cluster mass function and the halo definition 
from simulations has to be advanced. 

We have used our results for some exploration of the evolution
of the cluster population with redshift, and we showed how the
most massive clusters are confined to the low redshift range.
Ongoing and future surveys over a larger redshift range
will further tighten the mass function.
But here the situation is even more severe due to the 
uncertaities in scaling relations, since the
observational data on the relations are more sparse and less
precise at higher redshift. Therefore future surveys, have to provide not only
better samples but also a means to improve the mass-observable
scaling relations. In this respect the {\sf Euclid} mission 
with its capability of providing lensing masses for a 
large number of groups and clusters (e.g. Sartoris et al. 2016)
offers especially interesting prospects.

\begin{acknowledgements}
H.B. and G.C. acknowledge support from the DFG Transregio Program TR33
and the Munich Excellence Cluster ''Structure and Evolution of the Universe''.  
G.C. acknowledges support by the DLR under grant no. 50 OR 1405.
MF thanks Yasuo Tanaka at the Max-Planck-Institut fuer Extraterestische
Physik and the Alexander von Humboldt Stiftung for their support during his
stay in Garching. We like to thank the referee for useful 
comments which helped to make some points clearer.

\end{acknowledgements}

\end{document}